\pdfoutput=1

\documentclass[twocolumn,prd,amsmath,amssymb,superscriptaddress,nofootinbib]{revtex4}

\usepackage{aas_macros}

\usepackage{epsfig,psfrag,graphics,verbatim}
\usepackage{graphicx}% Include figure files
\usepackage{dcolumn}% Align table columns on decimal point
\usepackage{bm}% bold math
\usepackage{color}

\usepackage{hyperref}

\begin{document}

\title{Reply to Comment on ``Evidence for dark matter in the inner Milky Way''}

\author{Fabio Iocco}
\affiliation{ICTP South American Institute for Fundamental Research, and Instituto de F\'isica Te\'orica - Universidade Estadual Paulista (UNESP), Rua Dr. Bento Teobaldo Ferraz 271, 01140-070 S\~{a}o Paulo, SP Brazil}
\affiliation{Instituto de F\'isica Te\'orica UAM/CSIC, C/ Nicol\'as Cabrera 13-15, 28049 Cantoblanco, Madrid, Spain}

\author{Miguel Pato}
\affiliation{The Oskar Klein Centre for Cosmoparticle Physics, Department of Physics, Stockholm University, AlbaNova, SE-106 91 Stockholm, Sweden}
\affiliation{Physik-Department T30d, Technische Universit\"at M\"unchen, James-Franck-Stra\ss{}e, D-85748 Garching, Germany}

\author{Gianfranco Bertone}
\affiliation{GRAPPA Institute, University of Amsterdam, Science Park 904, 1090 GL Amsterdam, The Netherlands}

\date{\today}% It is always \today, today,
             %  but any date may be explicitly specified

\begin{abstract}
In a brief note posted recently, the authors of arXiv:1503.07813 raised some concerns on our arxiv:1502.03821, recently published in Nature Physics. We thank them for the interest in our work and respond here to their criticisms.
\end{abstract}
%\pacs{}% PACS, the Physics and Astronomy
%                             % Classification Scheme.
%\keywords{Suggested keywords}%Use showkeys class option if keyword
                              %display desired
\maketitle

%#####################

\par In a brief note posted on the arXiv, the authors of \cite{2015arXiv150307813M} raised concerns on our work ``Evidence for dark matter in the inner Milky Way'' \cite{2015NatPh..11..245I}, centred around three points: 1) our use of the term ``inner'' Milky Way; 2) the treatment of datasets and associated uncertainties; and 3) the novelty of our results. The text and supplementary information of our letter already contain a thorough discussion of all these points, but to avoid any confusion we shall briefly address them here again.

\par {\bf 1)} The authors criticise the use of the term ``inner'' to describe the Galaxy at Galactocentric radii smaller than $R_0\sim 8\,$kpc. Different communities use a different jargon, which in fact also varies through sub-communities, and with time. One could quarrel forever whether the Sun is in the ``outer'' Galaxy arguing, as the authors do, that the Solar circle encompasses most of the stellar mass, or in the ``inner'' Galaxy, as a cosmologist or dark matter phenomenologist may say, since the solar circle encompasses only about a few percent of the Milky Way virial mass. 

\par There is however no ambiguity in our letter. We have carefully and repeatedly described throughout our work what we mean by ``inner'' Galaxy and at which radii we obtain evidence for dark matter (see abstract, lines 11-13; paragraph 2, lines 10-12; paragraph 6, lines 12-14;  paragraph 7, lines 5-7; paragraph 8, lines 4-5). In particular, we have explicitly stated ``The analysis is restricted to Galactocentric radii $R > R_{cut} = 2.5\,$kpc, below which the orbits of the kinematic tracers are significantly non-circular.'' and then ``The discrepancy between observations and the expected contribution from baryons is evident above Galactocentric radii of $6-7\,$kpc.''. The ``innocent reader'' invoked by the authors could draw the same conclusion with a simple glance at Fig.~2 (bottom panel), in which the evidence is shown as a function of Galactocentric radius, without even reading the text.

\par {\bf 2)} The authors argue that ``One cannot simply combine [data from many different sources] to obtain a statistically meaningful estimate of the uncertainty'', and that ``Great care must be taken in doing this''. We agree on that, and in fact we have performed careful tests of our results against the impact of systematic errors, data selection, Galactic parameters and baryonic uncertainties. We encourage the authors to read the supplementary information of our work, where a very thorough discussion of these matters is presented. We are of course happy to discuss in a quantitative and rigorous way any technical concern the authors may have.

\par {\bf 3)} The discussion of the (perceived lack of) novelty of our results is grossly misleading. The most serious problem in Ref.~\cite{2015arXiv150307813M} is in the following statement: ``The necessity of DM only becomes clear beyond 6 kpc, where the rotation velocity exceeds the upper bracket of the baryonic contribution. This result is not new, and follows for any plausible value of the circular speed and Galactocentric distance of the Sun[10].''. The first sentence is the main result of our letter (cf.~paragraph 7, lines 5-7). The second sentence is factually wrong. 

\par The authors presumably confuse our approach with ``local'' measurements of the dark matter density, which are also discussed in the book of Binney and Tremaine (Ref.~10 in \cite{2015arXiv150307813M}), where an order-of-magnitude estimate of the local matter density is presented, under a set of simplifying assumptions, based on the dynamics of the Solar neighbourhood (see pp.~372-376, and especially Eq.~(4.279), in the 2008 edition of that book). We have acknowledged those and other measurements in our work (paragraph 1, lines 7-11 and 15-16), but in view of the large systematic and statistical uncertainties we opted for a different data-driven approach.

\par Determining the ``bracket'' due to all allowed morphologies of the visible component of the Milky Way, and showing that the inferred rotation curve of each such possible configuration is not sufficient to explain the kinematic data to high statistical significance, is the main point of our work. We have argued that such result -- obtained without assuming a specific dark matter profile -- provides in our opinion the first direct observational proof of dark matter inside the solar circle, because it is non-parametric, data-driven, robust against all uncertainties or nuisances and independent of the adopted morphology of the baryonic component of the Galaxy. 

%%%%%%%%%%%%%%%%%

\vspace{0.5cm}
{\it Note added.} McGaugh et al have replied to our three points above in the revised version of their comment \cite{2015arXiv150307813M}. We respond here to their remarks.

\par {\bf 1)} Regarding the use of the term ``inner'', we have nothing to add to our first reply above.

\par {\bf 2)} We have carefully quantified the uncertainty associated to each bulge, disc and gas across the Galaxy so that our results would be solid for any individual baryonic model. Then, as stated in our letter (see paragraph 5, lines 13-16), ``we do not attempt to account for the poorly understood systematics of each single baryonic model, but instead use the spread due to all morphological configurations as an estimate of the systematics on the baryonic contribution.''. This gave us a handle to test the robustness of our results against baryonic modelling.

\par {\bf 3)} McGaugh et al now point to Ref.~\cite{1988MNRAS.233..611S} to justify the lack of novelty of our work. That is certainly a well-known and very good article, which interestingly states the opposite of what McGaugh et al claim. In fact, the authors of Ref.~\cite{1988MNRAS.233..611S} write already in the abstract ``We present a mass model for the inner Galaxy which requires very little dark matter to account for the circular velocity at the Sun. [...] A dark halo may still be required to account for the rotation curve beyond the solar circle''. In the discussion the authors stress again that the need for dark matter arises only at larger radii: ``The rotation curve of our maximum disc model declines steadily beyond the solar circle [...]. Thus, as for other galaxies, the extended rotation curve will begin to reveal a mass discrepancy, which seems to require a massive non-luminous halo.''. 

\par The fact is that the inspection ``by-eye'' of figures like Fig.~2 in Ref.~\cite{1988MNRAS.233..611S} (and others in the literature), or approximate arguments on the radial force brought up by the authors in their first comment, can at most {\it suggest} the presence of dark matter at small Galactocentric radii. In our work we addressed the questions: How strong is the evidence? Is the discrepancy larger than the uncertainties? If so, down to what radius?

\par There are many ways to convince oneself of the presence of dark matter in the Milky Way at various radii, and it is certainly not surprising that it appears at some level in the data. In our letter we made a claim based on a specific technical point:  that current data are constraining enough to make the claim robust against statistical and systematic errors. We believe we have made this point clear with our letter and two replies, and we shall not continue the discussion on the arXiv. We encourage the authors to contact us directly should they have more doubts; we will be happy to respond.

%use bibtex
\bibliographystyle{apsrev}
\bibliography{reply}

%or output here the bbl file

\end{document}